# Mixed Effects Models are Sometimes Terrible[1]

Christopher Eager & Joseph Roy
*University of Illinois at Urbana Champaign*

**Abstract**

Mixed-effects models have emerged as the "gold standard" of statistical analysis in different sub-fields of linguistics (Baayen, Davidson & Bates, 2008; Johnson, 2009; Barr, et al, 2013; Gries, 2015). One problematic feature of these models is their failure to converge under maximal (or even near-maximal) random effects structures. Convergence tests are themselves different from version to version of statistical packages for mixed effects models and also differ across platforms (e.g. R, SPSS and SAS). The lack of convergence is relatively unaddressed in linguistics and when it is addressed has resulted in statistical practices (e.g. Jaeger, 2009; Gries, 2015; Bates, et al, 2015b;) that are premised on the idea that non-convergence is an indication that a random effects structure is over-specified (or not parsimonious), the *parsimonious convergence hypothesis* (PCH).

We test the PCH by running simulations in lme4 under two sets of assumptions for both a linear dependent variable and a binary dependent variable in order to assess the rate of non-convergence for both types of mixed effects models when a known maximal effect structure is used to generate the data (i.e. when non-convergence cannot be explained by random effects with zero variance). Data are simulated under two sets of conditions and for each simulated data set we fit the non-zero random effect structure used to simulate the data in lme4:

(1) With the simulation assumptions reported in Barr, et al. (2013) with 24 subjects and 24 Items and one binary predictor for both linear and logistic models.
(2) Across a range of imbalance in the data, with one binary and one three-level predictor, with a maximal random subject effect structure for 30-60 subjects (and no items).

In the case of (1), there are low rates of non-convergence for linear and logistic models. In (2), however, there is a much higher rate of non-convergence. Under the PCH, lack of convergence is treated as evidence against a more maximal random effects structure, but that result is not upheld with our simulations. We provide an alternative model, fully specified Bayesian models implemented in rstan (Stan Development Team, 2016; Carpenter, et al, in press) that removed the convergence problems almost entirely in simulations of the same conditions. These results indicate that when there is known non-zero variance for all slopes and intercepts, under realistic distributions of data and with moderate to severe imbalance, mixed effects models in lme4 have moderate to high non-convergence rates which can cause linguistic researchers to wrongfully exclude random effect terms.

## 1. Introduction

Over the last ten years, mixed effects models have become the gold standard of statistical analysis in linguistics and adjacent language sciences to replace ANOVA and regression models (linear and generalized linear) in the modeling of experimental, quasi-experimental and observational data. They provide several conveniences for data that come from repeated measures on participants and across items. They allow researchers to fully model all the sources of variability in the data, including repeated measures on the same subject and with the same experimental item (Baayen, Davidson & Bates, 2008; Barr, et al., 2013). These random effects, which extend traditional regression by adding alternative sources of variance, can also answer interesting sociolinguistic questions by accounting for subject mean differences (e.g. Drager & Hay, 2012) or by allowing social group variance to be modeled separately from residual variance (e.g. Eager, 2017a).

---

[1] The R code and simulation data are available for this paper:
https://github.com/jroy042/MixedEffectsAreSometimesTerrible The R package that can be used to implement the fully specified Bayesian model is available in Eager (2017b).

This study is part of a larger project by the 2nd author that aims to outline where mixed effects models can be problematic in practice (Roy, 2006, 2009, 2012, 2014; Roy & Levey, 2014; Kimball, et al. 2016). The *terribleness* alluded to in the title refers not to what these models are intended to do in linguistics (i.e. account for multiple observations on the same participant or item), but instead how these models behave with some kinds of actual data as well as practices linguistic researchers have developed in response to that behavior. The project has shown, with both real data (Kimball, et al. 2016) as well as the simulations discussed in this paper, that this *terribleness* can be alleviated with more constraints than standard mixed effects models implemented in lme4 provide: namely, by using a more fully specified Bayesian model.

To be very clear, we do not advocate using a simpler model (i.e. ordinary linear or logistic regression) in response to the problems we discuss when a researcher has repeated observations within subject or within item, nor are we suggesting that *lme4* be abandoned as a default implementation for these models. Instead we are advocating that fully specified Bayesian models that constrain assumptions on the fixed effects, the random effects and the covariance matrix might be better suited for categorical and imbalanced data. Specifically, in this paper, we demonstrate with a binary response that when there is moderate to severe imbalance in the data, a fully specified Bayesian model arrives at more stable results. As this paper is aimed at researchers who are familiar with the on-going discussions in linguistics on these models, but may not have formal training in statistics, we explain some of the terminology anticipated to be unfamiliar or notational variance anticipated to be confusing.

1.1 Mixed Effects Models in Linguistics

Over the last few years the authors have advised researchers on over 150 different projects across many sub-fields of linguistics as well as the adjacent language sciences that use mixed effects models, mostly with R and in lme4. The use of mixed effects models in linguistics typically follow recent advice on using these models (Bates, Davidson & Baayen, 2008; Barr, et al. 2013) in psycholinguistic experiments as an optimal control on type I error rates. This has expanded into other fields such as sociolinguistics and corpus linguistics (Johnson, 2009; Drager & Hay, 2012; Gries, 2015) as a means of accounting for individual differences at the subject or lexical item level in observational studies. In many of our conversations about mixed effects models, model convergence, as tested by lme4, is often a central issue of concern to the researcher. Convergence itself is not well understood by those who apply these models and even some who write software that is widely used in linguistics to implement these models[2]. The purpose of this specific study is to *pry open the black-box* (Hodges, 2014) of mixed-effect models with simulations based on real-world data sets as well as investigate simulations of imbalanced data to assess how well these models estimate parameters of interest.

The general form of a linear mixed model is formally (Demidenko, 2013: 45; Hodges, 2014: 5) defined as:

(1) $\boldsymbol{y = X\beta + Zu + \varepsilon}$

---

[2] E.g. Rbrul (Johnson, 2009), a program that uses lme4 but provides a menu-driven interface to help sociolinguists use mixed effects models, suppressed all convergence errors from lme4 from 2012-2013.

The $n \times 1$ vector, $y$, are the observed response variable. The design matrix of the fixed effects, $X$, represents a $n \times p$ matrix of all the predictors coded for each observation. The $p \times 1$ vector $\beta$ are the fixed effect coefficients. The random effect features form a $n \times q$ matrix, $Z$, and the $q \times 1$ vector $u$ are the random effect coefficients. The random effect structure is normally distributed with mean **0** and variance-covariance, $\Sigma$. The random coefficients $u$ are ancillary in the sense that they are not estimated directly in standard mixed effects model software: the covariance matrix $\Sigma$ is estimated and not $u$, the random intercepts and slopes, which can be estimated afterwards in several ways. The error, $\epsilon$, is normally distributed with a separate variance, $\sigma^2$.

This model can be extended to other kinds of response variables, where $g(y)$ replaces $y$ and $g$ is a one-to-one monotonic, differentiable function. For logistic regression, $g$ is the log-odds transformation of the probability of success, $p$, and we model:

$$(2)\ \ln \frac{p}{1-p} = X\beta + Zu$$

The probability that $y = 1$ (rather than 0) is represented by $p$. In language studies this can represent accuracy ($y = 1$ for accurate answers and $y = 0$ for inaccurate answers) or in eye-tracking when the participant's gaze is fixated on a particular region ($y = 1$ when fixated on the region of interest and $y = 0$ when they are not).

1.3 What is convergence?

Most researchers simply know convergence as the decorative warning messages angrily spit out by lmer or glmer when it does not seem the model is cooperating. Understanding what convergence tests do, however, is crucial for researchers applying mixed models. Convergence is a consequence of the iterative algorithm used to solve for $\Sigma$ and $\beta$ in (1) or (2). It consists of several tests that the results produced are statistically and computationally reliable. Usually this involves checking that $\Sigma$ is positive-semidefinite (i.e. that no linear transformation of the random effects has a negative variance estimate, which is an invalid estimate statistically, but possible computational result of the optimization algorithm), and that the difference between the second-to-last iteration and the last iteration in the optimization algorithm is smaller than a pre-determined tolerance (in lme4, this is done through checking the maximum gradient, with a tolerance of .002 in the current version, and in previous versions with a tolerance of .001). There are also checks on the fixed effects for statistical reasonableness, and checks which ensure there is enough data to estimate the proposed random effects structure (i.e. that the model doesn't have more parameters than observations).

Why is convergence not an issue in traditional linear regression or logistic regression? For linear regression, the calculation of the unknowns involves simple matrix operations that will not fail unless the predictors suffer from severe collinearity (and thus inverting the design matrix $X$ fails to produce a unique solution for the unknown predictors' effects). For logistic regression, there is an iterative algorithm, but it will not fail unless there is separation in the data (i.e. where too many cells have an observed $p$ of 1.0 or 0.0) or collinearity in the design matrix — in sociolinguistics this would be similar to knockout effects.

1.4 Parsimonious Convergence Hypothesis

In practice, the failure to converge is taken to be a mis-specification of the random effects structure. The desire to fit maximal models has resulted in ad-hoc practices, some reasonable[3] (but unattested in the formal statistical literature on these models) and some clearly **unreasonable**.

1. Use a PCA on covariance matrix to determine most meaningful slopes (Bates, et al., 2015b).
2. Reduce item random effect structure then reduce subject random effect structure until convergence (Jaeger, 2009).
3. Start with intercept only, use anova() to determine if a slope should be added or not. Stop when all slopes are not significant (Gries, 2015).
4. **Keep removing slopes randomly from item and subject random effect structure until convergence.**
5. **Suppress or ignore convergence errors.**

The Parsimonious Convergence Hypothesis (PCH) motivating these approaches *is that the failure of mixed effects model to converge is (most likely) due to the incorrect specification of the random effect structure*. In it is not clear from the statistical or applied statistical literature that convergence and parsimony are linked. The nature of mixed effects models and uncertainty around the appropriate random effects structure (as well as an inability to assess both random and fixed effects simultaneously in these models) leads to these ad-hoc practices which have been noted in other fields (Ryoo, 2011:599).

1.5 Experimental Approaches to Understanding Statistical Models

From Hodges (2014:xxxiii-xxxiv): *A lot of academics think mixed linear models are completely understood, when in fact they are still largely not understood...the new methods* [i.e. mixed effects models] *of the last three decades are so complex that it may never be possible to prove theorems about them. We can, however, make progress by approaching our black box methods in the same way our scientific colleagues approach nature's black-box methods, by prying them open gradually and indirectly if necessary.* Moreover, we would add that logistic mixed effects models are even less well understood[4].

---

[3] Some reasonable approaches do not produce the desired outcome, as we show.

[4] The interpretation of fixed effects are conditional on the random effects. This does not affect the interpretation of fixed effects with a linear mixed model, but does change the interpretation for logistic mixed models and can produce results that are even "*capable of confusing statisticians less familiar with these models*" (Molensberghs & Verbeke, 2005: 297).

Hodges (2016) continues to argue in favor of an experimental approach to assessing computational and statistical issues with mixed effects models which influences the approach to them in this paper. There are many examples of simulation studies in the theoretical and applied literature on mixed effects models. Agresti, Caffo, & Ohman-Strickland (2004) simulate two designs for a logistic mixed effects model in order to assess the effect of distribution specification on random effects: one within cluster random intercept and no other predictors and then one within-cluster random intercept with a binary fixed effect. Moineddin, Matheson and Glazier (2007), in order to do a power analysis for logistic mixed effects models, simulate a design of balanced data with one between cluster covariate. McCulloch & Nehaus (2011) use a design with two covariates and one level of random effect to assess how much misspecification of the variances affects both linear and logistic mixed effects models. Ryoo (2011), while using the a longitudinal study of student achievement scores as a basis for assessing model building techniques in linear mixed effects models with an actual complex design, still is using a mostly balanced design. None of these simulated designs, even Barr et al. (2013), come close to reflecting the complexity in designs we have seen in linguistic research. Near-balanced data is very much only possible when we have a well-designed and well-controlled experimental design that for many research programs is either not possible or is not practical to produce. Following the advice of Hodges (2014) to pry open the black box of mixed effects models, this study experimentally tests the PCH with near-balanced (simple) data and moderately to severely imbalanced (complex) data

## 2. Method and Data

This section describes the model setup and convergence criteria we use to assess our simulations. There is a lot of technical detail in this section necessary to reproduce and understand exactly what was simulated. We have attempted to make this accessible to non-mathematically inclined researchers who are reading as well as documenting why we made certain decisions. These simulations are meant to be more realistic than those mentioned in Section 1.4

2.1 Simple Mixed Effects Models

Following, Barr et al (2013) we simulate regressions with the following constraints: A total of 24 subjects and 12 or 24 items with at most 5% of the data removed on each simulation. Further, the simulation of one within subject effect that can be between item or not. As Bates et al. (2015b) discussed, this is not a typical design. Specifically, there is usually several within-subject effects that are modeled. Further, assuming the response is centered, one binary effect that shifts the response by .8 standard deviation is not typical of most linguistic experiments. This simulation, however, provides a control group for ours – it allows us to set a floor for non-convergence in balanced data sets. We use the terms "Gaussian" interchangeably with "Linear Mixed Effects Models" to represent a mixed effects model where the response is continuous and the error terms are (multivariate) normally distributed.

2.1.1 Linear Mixed Effects Models

The simulated model parameters for the fixed effects and random effects for Barr (et al., 2013) are provided below:

$$\beta_0 \sim uniform(-3,3); \; \beta_1 = 0 \; if \; H_0 \; is \; true; \beta_1 = 0.8 \; if \; H_0 \; is \; false;$$

| Subject RE | Item RE |
|---|---|
| $\sigma_{intercept_{subject}} \sim \sqrt{uniform(0,3)};$ <br> $\sigma\_slope_{subject} \sim \sqrt{uniform(0,3)}$ <br> $\rho_{subject} \sim uniform(-0.8, 0.8)$ | $\sigma_{intercept_{item}} \sim \sqrt{uniform(0,3)};$ <br> $\sigma\_slope_{item} \sim \sqrt{uniform(0,3)}$ <br> $\rho_{item} \sim uniform(-0.8, 0.8)$ |

The intercept, $\beta_0$, and the within subject fixed effect, $\beta_1$, are both described above, with only the intercept varying between -3 and 3. Only half the simulated data sets have a within-item effect (which would include the parameters highlighted in blue).

2.1.2 Logistic Mixed Effects Models

The logistic model parameters cannot just be copied from the linear one above; we need to add a few more constraints. The initial design is the same, but we must decrease the amount of variability, especially since we are concerned with the likelihood of convergence. The distributions these model parameters were drawn from were designed to make true log-odds values for logistic responses which were outside of [-5, 5] rare (henceforth, "extreme values"). The reason for this is that there is not enough information in the generated datasets (or in many real datasets; see Kimball et al. 2016) to make meaningful distinctions between, say, a log-odds of 5 and a log-odds of 6. This difference corresponds to the difference between Bernoulli probabilities of 0.9933 and 0.9975, which would not be recoverable from the dataset given that each speaker has on average 12 or 24 observations. We restricted the variance components in order to reduce the average percentage of extreme values to less than 1% of the data (for the simulated models reported, the mean percent of extreme values is 0.2%). The new random effect structure is defined below:

| Subject RE | Item RE |
|---|---|
| $\sigma_{intercept_{subject}} \sim uniform(0,1)$ <br> $\sigma\_slope_{subject} \sim uniform(0, .75)$ <br> $\rho_{subject} \sim uniform(-0.9, 0.9)$ | $\sigma_{intercept_{item}} \sim uniform(0,1)$ <br> $\sigma\_slope_{item} \sim uniform(0, 0.5)$ <br> $\rho_{item} \sim uniform(-0.9, 0.9)$ |

2.1.4 Number of simulations

To replicate the procedure used by Barr and his colleagues, we began with 80,000 total simulations for the simple linear situation (10,000 across each condition), but decreased this for the simple logistic regression situation (with 2500 per condition). The reason for doing this was practical: the 20,000 simple logistic simulations required the same amount of time as the 80,000 simulations – about a week. Moreover, it is not clear that the large number of simulations are necessary for what we want to assess. Barr and colleagues chose 800,000 total simulations for their results, but there is not a statistical reason, that we know of, for such a high number of simulations. We choose a smaller number for similar practical reasons for the complex models described in Section 2.2.

2.2 Complex Linear and Logistic Regression

One of the drawbacks to simulation studies is that they can often fail to represent realistic datasets. For example, Barr et al. (2013) simulate only linear models with a binary factor as the sole fixed effect, balanced across subjects and items, with between 0% and 5% (determined from a random uniform variable) of data simulated as "missing." Even in controlled experiments, this type of simple fixed effects structure is highly unlikely, and in observational studies in linguistics, such a high degree of balance among the subjects is hardly ever achievable due to the nature of the data. It is important to establish how this imbalance affects linear and logistic data differently.

In this study, both logistic and Gaussian mixed effects models are simulated for a fictional subjects-only observational design. The fixed effects are one binary factor $x_1$ and one three-leveled factor $x_2$ (whose levels will be referred to alphabetically: $x_1 \in \{a,b\}$, $x_2 \in \{a,b,c\}$), and varying degrees of imbalance were simulated for the number of subjects, the number of observations per subject, and the relative frequency with which the levels of the factors occur. Letting $S$ be the number of subjects for a model, $\lambda$ be the mean number of observations per subject, $n_s$ be the number of observations for subject $s$, $n$ be the total number of observations, and $p(x_1)$ and $p(x_2)$ be the relative frequencies of the levels of the factors, each dataset was generated per the following procedure:

$$S \sim Uniform(\{30,\ldots,60\}), \qquad \lambda \sim Uniform(20,30),$$

$$n_s \sim \max\{Poisson(\lambda), 1\}, \qquad n = \sum_{s=1}^{S} n_s,$$

$$p(x_1) \sim Dirichlet(1,1), \qquad p(x_2) \sim Dirichlet(1,1,1).$$

That is, for each model, the number of subjects was chosen uniformly between 30 and 60, the mean number of observations per subject was uniformly distributed on [20, 30], and the number of observations per subject was drawn from a Poisson distribution with that mean but ensuring that the minimum was 1 observation per subject (which should nearly never be necessary). The relative frequency of the levels of $x_1$ and $x_2$ were drawn from standard Dirichlet distributions, and then, regardless of the number of observations a subject had, each observation's values for $x_1$ and $x_2$ were determined given the probabilities generated by the Dirichlet distributions for the dataset (the Dirichlet distribution is a probability distribution on a vector of random variables each on (0,1), and whose sum is always 1; this makes it the ideal choice for generating random distributional frequencies; when the parameters of the distribution are all set to 1, then all possible distributional vectors are equally likely).

*2.2.1 Balance*

To measure balance in the dataset, our goal was to create a measure on (0,1] where 1 is a perfectly balanced dataset, and the measure moves toward zero as the dataset becomes more imbalanced. Balance in this case means the same number of observations produced for every level of a predictor in every subject's data. Researchers not familiar with corpus or sociolinguistic data may not see the reason for such a measure (or even discussion of imbalanced data), but most data sets that arise from language production (via any modality) have some imbalance. Many have severe imbalance. The only way to really assess these models statistically and account for imbalance is to have a measure that allows us to capture imbalance in a data set for a given set of predictors in reproducible and principled way.

First, a contingency table was created for $Subject * x_1 * x_2$; that is, a table with a count of the number of observations for each combination of $x_1$ and $x_2$ for each subject, containing $\mathcal{C} = \mathcal{S} * 6$ cells. The number of non-empty cells (count greater than zero) was then determined, denoted as $\mathcal{C}^*$. The mean count for the non-empty cells was then calculated as $n^* = n/\mathcal{C}^*$. For each non-empty cell $c_i^*, i \in \{1, \ldots, \mathcal{C}^*\}$, an imbalance ratio was calculated as $r_{c_i^*} = \max\{n_{c_i^*}/n^*, n^*/n_{c_i^*}\}$. The mean of these ratios was then multiplied by the proportion of empty cells to obtain an overall ratio; that is,

$$r = \frac{\mathcal{C}}{(\mathcal{C}^*)^2} \sum_{i=1}^{\mathcal{C}^*} r_{c_i^*}.$$

The overall imbalance ratio $r$ is bounded by $[1,\infty)$, with higher values indicating more imbalance. To create a measure on (0,1] where higher numbers indicate more balance in the dataset, the ratio was transformed to create a balance measure $\mathcal{B}$:

$$\mathcal{B} = 2\left[1 - \frac{r}{1+r}\right].$$

If the data are perfectly balanced (i.e. if all cells in the contingency table have the same count), then we have:

$$\mathcal{C}^* = \mathcal{C},$$

$$n^* = \frac{n}{\mathcal{C}^*} = \frac{n}{\mathcal{C}} = n_{c_i^*} \text{ for all } i \in \{1, \ldots, \mathcal{C}\},$$

$$r_{c_i^*} = \max\{n_{c_i^*}/n^*, n^*/n_{c_i^*}\} = \max\{1,1\} = 1 \text{ for all } i \in \{1, \ldots, \mathcal{C}\},$$

$$r = \frac{\mathcal{C}}{(\mathcal{C}^*)^2} \sum_{i=1}^{\mathcal{C}^*} r_{c_i^*} = \frac{1}{\mathcal{C}} \sum_{i=1}^{\mathcal{C}} 1 = 1,$$

$$\mathcal{B} = 2\left[1 - \frac{r}{1+r}\right] = 2\left[1 - \frac{1}{2}\right] = 1.$$

As the data become more imbalanced, $\mathcal{B} \to 0$, but cannot have an actual value of 0. This balance measure helps us to establish how imbalance affects the accuracy of the parameter estimates and likelihood of non-convergence for the simulations. As we move forward with this research, we anticipate refining this imbalance measure for other designs as well as extending a similar measure for coverage with a continuous predictor. Finally, it is also important to state that "imbalanced" data is not equivalent to a "badly designed study", especially with observational data where you cannot be certain what linguistic contexts will produce themselves in an hour of interview data, interaction data, letters or whatever unstructured data you have.

*2.2.2 Generation of the true model*

For the generation of the true model (and for the regressions), sum contrasts were used for the factors. The fixed effects model matrix $X$ thus contains four columns coded as follows:

$$X_0 \ (intercept) = 1, \qquad X_1 = \begin{cases} +1 & x_1 = a \\ -1 & x_1 = b \end{cases},$$

$$X_2 = \begin{cases} +1 & x_2 = a \\ 0 & x_2 = b \\ -1 & x_2 = c \end{cases}, \qquad X_3 = \begin{cases} 0 & x_2 = a \\ +1 & x_2 = b \\ -1 & x_2 = c \end{cases}.$$

A mixed effects model with a full-rank random effects variance-covariance matrix thus requires the specification of four fixed effects coefficients, a four by four covariance matrix for the subject random effects, intercepts and slopes for each subject, and, for Gaussian models, the error variance and random errors. The distributions these model parameters were drawn from were designed to make extreme (outside of [-5,5]) true log-odds values for logistic responses rare. The following true model distributions were found to produce very few extreme values, with 98% of the models generated having fewer than 1% extreme values (the same distributions were also used for linear models, as in this case the response will be mean-centered and scaled prior to analysis, and the range is not as important; $MN$ is the multivariate normal distribution):

$$\beta_0 \sim U(-2,2), \qquad \beta_j \sim U(-1,1) \ for \ j \in \{1,2,3\},$$

$$\sigma_{S0} \sim U(0,1), \qquad \sigma_{Sj} \sim U(0,0.5) \ for \ j \in \{1,2,3\}, \qquad \Omega_S \sim LKJ(1),$$

$$\Sigma_S = \text{diag}(\sigma_S)\Omega_S \text{diag}(\sigma_S)^T, \qquad \gamma_S \sim MN(\mathbf{0}, \Sigma_S),$$

$$[For \ Gaussian] \ \sigma_\epsilon \sim U(0,1), \qquad \epsilon \sim N(0,\sigma_\epsilon^2), \qquad y = X\beta + Z\gamma + \epsilon,$$

$$[For \ Logistic] \ \ln\left(\frac{p}{1-p}\right) = X\beta + Z\gamma, \qquad y \sim Bernoulli(p).$$

As can be seen above, rather than define the random effects covariance matrix $\Sigma_S$ directly, the random effects correlation matrix $\Omega_S$ and the standard deviations for each effect $\sigma_{Sj}$ were sampled independently, with the covariance matrix being a by-product of the two. The LKJ distribution with parameter $\eta = 1$ is uniform over the space of all possible correlation matrices (Stan Development Team 2016). The fixed effect for the intercept and the random effects variance for the subject intercepts were given wider distributions than the slopes, as this is often the case in real datasets.

### 2.2.3 Fitting procedure for lme4

For linear mixed models models, the response $y$ was mean-centered and scaled prior to analysis, as this increases the likelihood of convergence (Bates et al. 2015a). For both Gaussian and logistic models, the following lme4 formula was used, representing the maximal random effects structure as defined by Barr et al. (2013):

$$y \sim x_1 + x_2 + (1 + x_1 + x_2 \mid subject).$$

For the linear mixed models, the function lmer was used (REML) and for logistic models, the function glmer was used with family = binomial (ML).

### 2.2.4 Fitting procedure for Stan

For the Bayesian models fit in Stan, weakly informative priors were used, extending those used in Gelman et al. (2008). The term *prior* can be confusing for researchers unfamiliar with Bayesian approach modeling data. From an intuitive perspective, the model we describe below as a "fully specified Bayesian mixed effects model" is different from a linear mixed effects model in how the parameters are constrained. Interested readers can refer to McElreath (2016) or Kruschke (2014) for book-length introductions to Bayesian statistical methods or to Nicenboim & Vasishth (2016) for an article-length introduction with a linguistic focus. Essentially, the Bayesian approach allows us to make reasonable constraints on the fixed and random effects that allow us to estimate parameters more readily. These constraints, however, are not binding on the posterior estimates – if there is evidence in the data that the starting constraints are incorrect, the posterior estimates will be driven by the data. Sometimes these constraints map onto real world considerations of the research and other times these constraints are computationally required. There are broader benefits to this class of models (including being able to move beyond a Null-Hypothesis Significance Testing (NHST) approach to statistical modeling in data analysis) but these are beyond the scope of this paper.

For linear mixed models, the response was mean-centered and scaled prior to analysis. The following priors were used, where $MN$ is the multivariate normal distribution parameterized with $\boldsymbol{\mu}$ and $\Sigma$, $HN$ is the half-normal distribution parameterized with a scale parameter, and $N$ is the normal distribution parameterized in terms of $\sigma$ rather than $\sigma^2$:

$$\beta_j \sim N(0,2) \; for \; j \in \{0,1,2,3\},$$

$$\sigma_{Sj} \sim HN(0,1) \; for \; j \in \{0,1,2,3\}, \quad \Omega_S \sim LKJ(2),$$

$$\Sigma_S = \text{diag}(\sigma_S)\Omega_S\text{diag}(\sigma_S)^T, \quad \gamma_S \sim MN(\mathbf{0},\Sigma_S),$$

$$[Gaussian] \quad \sigma_\epsilon \sim HN\left(0,\frac{1}{2}\right), \quad y \sim N(X\beta + Z\gamma, \sigma_\epsilon),$$

$$[Logistic] \quad \ln\left(\frac{p}{1-p}\right) = X\beta + Z\gamma, \quad y \sim Bernoulli(p).$$

The $LKJ(2)$ prior is conceptually similar to placing a $2 * Beta(2,2) - 1$ prior on a single correlation coefficient, but in the multivariate case the mode at zero represents the identity matrix (Stan Development Team 2016). In the Stan code in Appendix B, the parameters are sampled on unit scale or in terms of Cholesky factors (as applicable) and then reparameterized, following the guidelines in the Stan user manual (Stan Development Team 2016). For each Stan model, three chains were run with 1000 warmup iterations and 1000 post-warmup iterations, with the target pseudo-acceptance rate $\delta$ set to 0.99, and with the initial values of the parameters in each chain being randomly chosen by the Stan algorithm. A model was labeled as having converged if there were no divergent transitions post-warmup and if the Gelman-Rubin $\hat{R}$ statistic was less than 1.1 for all parameters, and as unconverged otherwise.

*2.2.5 Simulation*

A total of $M = 5000$ datasets were simulated, with a Gaussian model generated for 2500 and a logistic model generated for the other 2500. For each model, two regressions were fit, one with lme4 and one with Stan, resulting in 10,000 regressions total. For each regression, the following information was recorded. We focus on the most relevant results in this paper to convergence and researcher behavior (the other parameters can be very interesting from a statistically computational point of view, but tend not to be relevant to practitioners):

- model family (Gaussian or logistic)
- regression type (lme4 or Stan)
- balance $\mathcal{B} \in (0,1]$
- the true model parameters $\beta, \sigma_S, \Omega_S$ and, for Gaussian models, $\sigma_\epsilon$ (with Gaussian parameters expressed on mean-centered unit scale)
- the regression estimates for the model parameters and the squared error for each
- for lme4 fits, whether the random effects Cholesky factor estimate was rank-deficient
- whether or not the model converged
- the smallest random effects variance in the true model
- the sum of the squares of the lower triangle of the correlation matrix (total correlation)

2.3 Convergence

Regression convergence in lme4 was assessed through the messages returned as part of the fitted object. If a warning message indicated an unidentifiable model, a degenerate Hessian matrix, or a maximum change in the gradient between the final two iterations greater than or equal to 0.01, the model was labeled as unconverged; otherwise, it was labeled as converged.

The tolerance we use is 5 times higher than the lme4 default (.002). This lme4 default itself has been doubled recently from .001 to .002. The reason for the tolerance used in this study is simply than most of the logistic models have difficulty meeting the .002 tolerance and our results for the complex logistic models would have had only a handful of converged models. A practical consideration in increasing the number of iterations was time: the 2500 simulations took a week to run on a high-speed, high-RAM computer.   The appropriate tolerance level is not a mathematically straightforward task to evaluate (see Demidenko, 2013: 678-679 for a discussion). The authors intuition for mixed effects models is that issues with the max|grad| tolerance is the least problematic of convergence errors and, in practice, many researchers tend to ignore[5] this error if there are no others.

## 3. Results

Table 1. Overall rates of non-convergence for simple and complex models

| MEM Type | Non-Convergence *lme4* | Non-Convergence *RStan* | Number of Simulations |
|---|---|---|---|
| **Simple Linear** | 0 % | .009 % | 80,000 |
| **Simple Logistic** | 7 % | .002 % | 20,000 |
| **Complex Linear** | 14 % | 3 % | 2,500 |
| **Complex Logistic** | 82 % | <.001 % | 2,500 |

In Table 1, the overall rates of non-convergence are presented. Comparing the simple linear model in both lme4 and RStan implementation, there does not seem to be much of a practical difference: almost all simulated data sets converge for both (0% and .009% respectively). This is not true moving to the simple logistic simulations.  There is a slight, but noticeable, amount of non-convergence in lme4 simple logistic regressions (7% versus the .002% for RStan).  In the complex situation, both linear and logistic models have a substantial increase in the amount of non-convergence for lme4.  For the linear model, 14% of the simulations fail to converge in lme4 while only 3% fail to converge in RStan. For the logistic model, 82% fail to converge for lme4 while much less than 1% fail to converge in RStan.

We return to these in Section 4, but to explore what might condition non-convergence, we dig deeper into our data.

3.1 Imbalance, Minimum Variance, Sample Size and Convergence.

We fit a generalized additive model (Wood, 2006) to the likelihood of convergence with family (linear or logistic), a smooth for minimum variance for the random effects and a separate smooth for the balance ratio. The full table of results are included in the Appendix B. The visualization of the model estimates, however, are the most important for this paper and provide an accessible way to discuss the impact of each covariate on the likelihood of convergence.

---

[5] Ignoring this error is not appropriate: a large max|grad| value could indicate that the estimates need more iterations to converge or that the algorithm is having difficulty locating a solution (i.e. headed in the wrong direction).

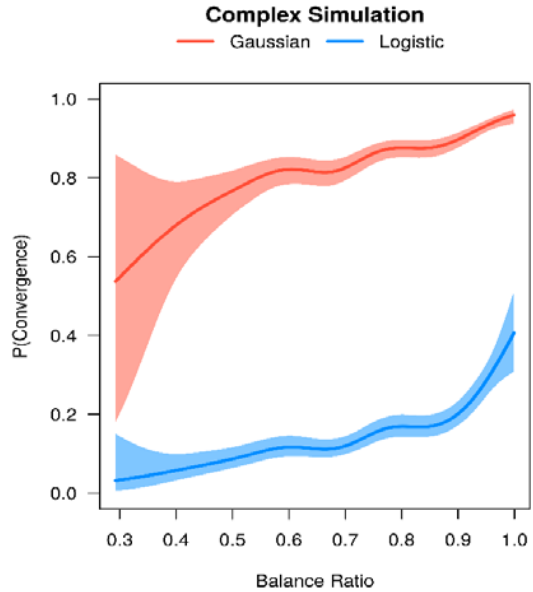

Figure 1: Likelihood of Convergence across Balance

In Figure 1, the predicted probability of convergence in the simulations is displayed across the range of the balance ratio. The solid red and blue lines represent the estimated predicted convergence probability from the model. The shaded area represents 95% predicted intervals from the model. There is a distinction between linear and logistic models that is seen in the remainder of the results (i.e. complex logistic models are much less likely to converge.) The likelihood of convergence, however, increases as the data sets become more balanced.

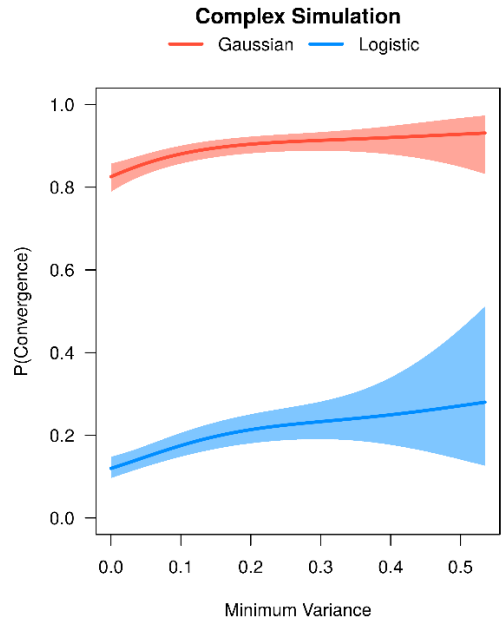

Figure 2: Likelihood of Convergence across Minimum Variance

In Figure 2, the predicted probability of convergence across the minimum variance in the random effects structure is displayed. For the complex linear models, when the minimum variance is close to 0.0, the predicted rate of convergence is still close to .80 while that rate rises to almost 1.0 as the minimum variance increases. The same direction of effect is seen in the complex logistic models with an increase from .16 to almost .30 at the maximum value for minimum variance in our data.

3.3 Accuracy of Fixed Effects in Converged Models

Again, we fit a generalized additive model (Wood, 2006) to the predicted error for type of parameter, type of mixed effects model and balance ratio. The predicted error for linear and logistic error are modeled within one gam. The full results are presented in Appendix B.

In Figure 3, the estimated effects for predicted error in complex linear models in both lme4 and RStan estimation is shown, but only for those that converged. For what researchers are most interested in, the fixed effects and random effect variances, we find almost no difference in the converged estimation between lme4 and RStan. In Figure 4, the estimated effects for predicted error in complex logistic models in both lme4 and RStan estimation is shown. The first difference between this graph and Figure 3, is that the fixed effects are more prone to error with lme4 than RStan. This differentiation between the two models continues for the random effect variation and correlation structures. The correlation and variance difference, as stated above, are not usually relevant to a research question, but the fixed effects are absolutely central to most research that use mixed effects models. The largest error is in the intercept (or, under sum contrasts, the corrected mean) which is not as problematic, but in both the binary effect and the three-level effect, parameter estimates have a higher error in lme4 than RStan.

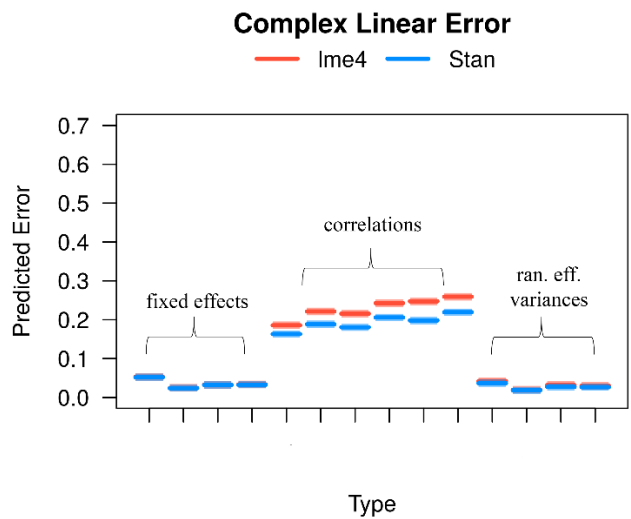

Figure 3: Predicted Error for Complex Linear Mixed Models that Converged

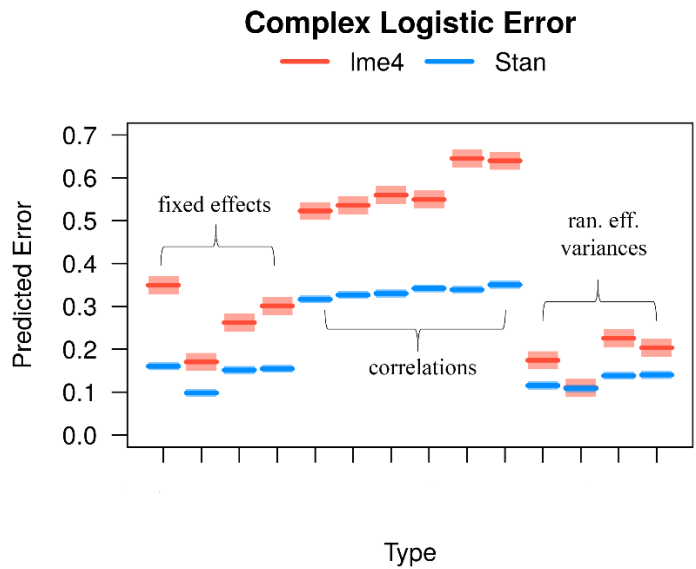

Figure 4: Predicted Error for Complex Logistic Models that Converged

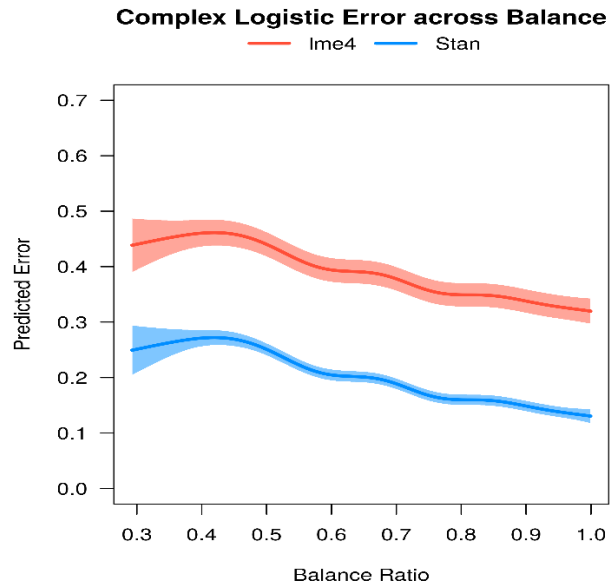

Figure 5: Predicted Error for Complex Logistic Models that Converged across Balance

From Figure 5, the difference between lme4 and RStan seen for non-convergence is maintained for predicted error in the complex logistic models. The predicted error, however, in both cases, decreases as balance increases with completely balanced data sets (at 1.0) having the least amount of predicted error in the parameter estimates for both lme4 and RStan.

3.4 Rank of Correlation Matrix

The correlation matrix may be of little practical use to researchers, but Bates and colleagues (2015b) propose using it to determine the optimal random effect structure. To determine whether or not an lme4 estimate of the random effects covariance matrix $\Sigma_S$ was rank-deficient, we performed principal component analysis on the Cholesky factor of the matrix. The matrix was coded as rank-deficient if fewer than four principal components (the dimension of the matrix) were required to cumulatively account for 100% of the variance, and full-rank otherwise (Bates et al. 2015b). This portion of the analysis was only performed on lme4 regressions which converged (n = 1041), as the technique is not reliable in unconverged regressions. As shown in Table 2, in more than half of the Gaussian regressions and nearly all of the logistic regressions, the lme4 estimate of the covariance matrix was rank-deficient even though the true model is always full-rank.

Table 2. Rank-deficiency in converged lme4 estimates of $\Sigma_S$ by model family

| Rank | Gaussian | Logistic |
| --- | --- | --- |
| Full | 395 | 4 |
| Deficient | 469 | 173 |

Bates et al. (2015b) claim that this diagnostic indicates that there are zero-variance components or components which are perfectly correlated (i.e. model misspecification). As shown in Figure 6, small variance components and high correlations do increase the likelihood that lme4 will estimate the matrix as rank-deficient, but rank-deficient lme4 estimates also clearly arise when all variance components are non-zero, and when correlations are low.

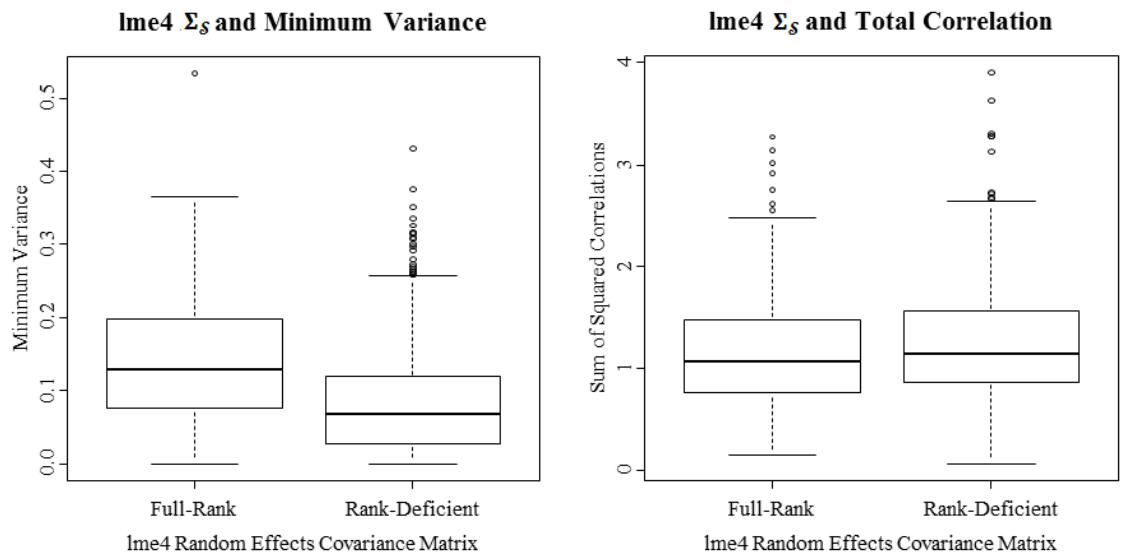

Figure 6. Boxplots of minimum variance (left panel) and total correlation (right panel) by converged lme4 estimation of the rank of $\Sigma_S$. Furthermore, as shown in Figure 7, more imbalanced datasets are also more likely to result in rank-deficient lme4 estimates of $\Sigma_S$.

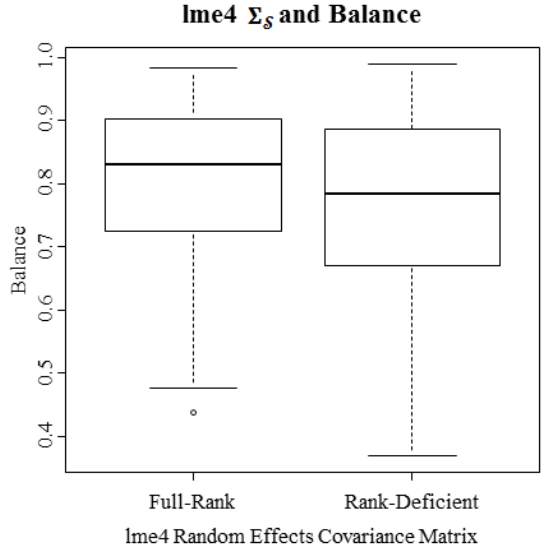

Figure 7. Boxplot of dataset balance by converged lme4 estimation of the rank of $\Sigma_S$

A logistic regression with rank estimation as the response (full-rank = 0, rank-deficient = 1) shows that all three of these predictors (minimum variance, balance, and total correlation) have statistically significant effects in the data in addition to logistic models being more likely to be found rank-deficient (Table 3).

Table 3. Logistic regression on the rank of converged lme4 estimates of $\Sigma_S$ (n = 1041)

| Fixed Effect | Estimate | SE | z | p |
| --- | --- | --- | --- | --- |
| Intercept | 2.474 | 0.277 | 8.918 | < .001 |
| Family, Gaussian | -2.328 | 0.278 | -8.360 | < .001 |
| Minimum Variance (Scaled) | -1.009 | 0.091 | -11.110 | < .001 |
| Balance (Scaled) | -0.555 | 0.081 | -6.847 | < .001 |
| Total Squared Correlation (Scaled) | 0.217 | 0.078 | 2.768 | .006 |

## 4. Discussion

It is clear both in linear and logistic models with complex imbalance in the data, that lme4 may not be the most appropriate implementation of mixed effects models. Further, even when the models converge, lme4 seems to provide more error in each model than RStan, although for linear models this is only manifested in the correlation estimates. Below, we explore possible explanations for differences we have seen with respect to convergence and error in estimating the parameters for linear and logistic mixed effects models across both lme4 and RStan.

4.1 Logistic versus Linear Mixed Models

The most striking difference, regardless of using RStan or lme4 to estimate the model, is the difference between linear and logistic mixed effects models. The rate of convergence for lme4 is terrible with the complex logistic data. The parameter estimates in the complex logistic data set are also similarly worse than RStan for lme4. Why is this not seen in linear mixed effects models? From an informational perspective, there simply is not a lot of information in each observation for a logistic model – the binary response (0 or 1) does not give as much information about the probability $p$ as an observation in a linear model gives about the true $y$. It becomes harder to find solutions (iteratively) for the logistic model and the statistical properties of the fixed effects estimates themselves differ in very meaningful ways from those produced in a linear mixed effects model (Molensberghs & Verbeke, 2005:255-259).

4.2. A fully specified Bayesian Mixed Effects Model

Assessing the underlying estimation algorithm in lme4 is beyond the backgrounds of both authors, but we have some ideas that do not point to a failure in lme4, but rather to the benefit of using a bigger model (i.e. with more constraints). The weakly informative priors put forward in Section 2 make reasonable assumptions about the data that help locate optimal parameter estimates when there is missing information (i.e. imbalance) or little noise in the random effects. Without these constraints, the lme4 algorithm tends to have difficulty locating estimates (especially in the imbalanced data, but still can be seen with the near-balanced data simulations). An imperfect analogy would be that if you misplaced your keys in your house, you probably should start looking around your house and not fly 12 hours away and start looking there. We suspect that in some circumstances the lme4 algorithm makes the 12-hour flight instead. This should not be taken to mean that lme4 is a "bad algorithm"— we do not know how other procedures would perform to estimate mixed effects models (such as nlme, SPSS or SAS implementations) and in many cases, as we have shown, it works well. We also fully recognize the difficulty in implementing these models and the computational skill required to do so. It is only through the development of lme4 and its widespread use that mixed effects models have become available in linguistics and the broader language sciences.

4.2 Power and Mixed Effects Models

We have ignored the issue of sample size in this study to control for the effect of balance and minimum variance. Sample size and statistical power, obviously are important to researchers. Statistical power, intuitively, is the likelihood of detecting a true statistically significant result if there is one present in the data. Underpowered studies (i.e. those without enough observations to reliably detect real effects) are very problematic because they increase the likelihood of Type M (magnitude) and Type S (sign) errors (Gelman & Carin, 2014). This means that having a smaller than necessary sample size makes overestimating effects and estimating them in the wrong direction more likely. The problem, of course, with doing power analysis outside of a small toy design is fixing reasonable windows on the parameters (e.g. what in 2.2.2 would be reasonable values for an analysis if there were more predictors, a covariate and items thrown in as well). Our results with the small window of sample size we manipulated, from 30 to 60 subjects, do not indicate that likelihood of convergence within this window is affected by sample size. We do not anticipate that finding to hold with larger windows of sample sizes and more complex designs. More broadly, there is some evidence in the applied statistical literature that many linguistic studies with a mixed effects model design (linear or other otherwise) are woefully underpowered. Ryoo (2011) found that when both random and fixed structure is in doubt, for his simulated data, that only at about 200 subjects were the analyst able to correctly identify the appropriate model for a linear mixed effects model. Moineddin, Matheson and Glazier (2007) found that in a very simple design with one within cluster covariate, that at least 50 clusters with 50 within cluster observations were needed in a logistic mixed effects model to generate reliable estimates. We expect that number, for the logistic mixed effects model, to increase as complexity is added to the design.

## 5. Conclusions

With regard to the behavior of mixed effects models themselves, we have more questions at the end of this set of experiments than answers. While we have shown rather conclusively that mixed effects models as currently implemented by lme4 have problems with imbalanced data and binary responses, there are still several issues to be addressed. We do not know what the likelihood of convergence would be for models with true zero random effects added into the model or if the RStan implementation we present here would perform better. We also do not know what adding more subjects would do to this and while we sample on an interval of 30 to 60. We anticipate some readers thinking that if we had "enough" subjects for a logistic mixed effects model, there would be a higher likelihood of convergence. Our experience is that larger data sets with moderate to severe levels of imbalance would have more problems converging than those presented in this paper. There is a clear distinction between a binary response (conditioned on the binomial distribution) and a continuous response (conditioned on the Gaussian distribution), in probability of convergence and predicted error between lme4 and RStan, with the Bayesian model in RStan outperforming lme4 for logistic and imbalanced data. We strongly suspect that this distinction will be maintained with more diverse and differently imbalanced data sets. The asymptotic approaches with unrealistic simulations to determine statistical model structure in much of the academic statistical literature fail to capture the diversity (and limits on N) present in many data sets that these models are used on. The experimental approach to statistics advocated by Hodges (2014; 2016), specifically for mixed effects models, offers a way forward of identifying and hopefully addressing these issues in mixed effects models.

Finally, we recognize the reaction of some linguists to advocating Bayesian models is something along the lines of "*not another damn model*." Over the last ten years, there has been much *Sturm und Drang* over which statistical model is most appropriate for data across the language sciences and many researchers feel overwhelmed with staying on top of new developments in their field, along with the seemingly ever-changing statistical requirements being asked from reviewers. Furthermore, there is the often unstated, but certainly present, feeling among researchers that those advocating newer statistical models are trying to attack established results via the back door of quibbling over mathematical nuances that only a statistician truly understands. Our response to this is simply to acknowledge that established results may be wrong and may even be wrong because of the statistical tools used at the time they were established. Much of the current replication crisis, however, is often not an artefact of insufficiently advanced statistical models, but what researchers do to the data before submitting it to a statistical model (Simmons, Nelson, & Simonsohn, 2011) or ignoring the *multiverse* of possible operationalizations for a given data set and research question (Gelman and Loken, 2013; Steegan, et al., 2016). Further, as more computational power becomes widely available and more statistical advances occur, we expect there to be even more statistical models over the next decade to appear in the linguistic literature. The driving force behind any of these models past, present or future, should be how they help illuminate the way language at all its levels, in all its modalities and in all its social environments works. The fully specified Bayesian models we present here in this paper reflects a model that is motivated not by overturning established results, but by answering newer questions and opening other research programs: traditionally linguistics studies have focused on statistically establishing group mean differences, but in more and more complex situations, we can explore differences in group variance[6] within a Bayesian framework (Eager, 2017a) and unaccounted for groups in individual differences (Drager & Hay, 2012).

**References**

Agresti, A., Caffo, B., & Ohman-Strickland, P. (2004). Examples in which misspecification of a random effects distribution reduces efficiency, and possible remedies. *Computational Statistics & Data Analysis*, *47*(3), 639-653.

Baayen, R. H., Davidson, D. J., & Bates, D. M. (2008). Mixed-effects modeling with crossed random effects for subjects and items. *Journal of memory and language*, *59*(4), 390-412.

Bates, D., Maechler, M., Bolker, B., & Walker, S. (2015a). Fitting Linear Mixed-Effects Models Using lme4. *Journal of Statistical Software* 67(1): 1-48.

Barr, D. J., Levy, R., Scheepers, C., & Tily, H. J. (2013). Random effects structure for confirmatory hypothesis testing: Keep it maximal. *Journal of memory and language*, *68*(3), 255-278.

Bates, D.M., Kliegl, R., Vasishth, S., and Baayen, H. (2015b). *Parsimonious mixed models*. Journal of Memory and Language (submitted). https://arxiv.org/pdf/1506.04967

Carpenter, B., Gelman, A., Hoffman, M., Lee, D., Goodrich, B., Betancourt, M., ... & Riddell, A. (2016). Stan: A probabilistic programming language. *Journal of Statistical Software*, in press.

Demidenko, E. (2013). *Mixed models: theory and applications with R*. John Wiley & Sons.

Drager, K., & Hay, J. (2012). Exploiting random intercepts: Two case studies in sociophonetics. *Language Variation and Change*, *24*(01), 59-78.

Eager, Christopher. (2017a). Modeling Complex Random Effects Structures in Sociolinguistics. *Linguistics Society of America 2017 Annual Meeting*, Austin (January 5-8).

----(2017b). *Bmers: Bayesiean mixed effects models with stan*. R package available on github: https://github.com/cdeager/bmers

Gelman, A., & Carlin, J. (2014). Beyond power calculations assessing type s (sign) and type m (magnitude) errors. *Perspectives on Psychological Science*, *9*(6), 641-651.

Gelman, A., Jakulin, A., Pittau, M.G., and Su, Y.S. (2008). A weakly informative default prior distribution for logistic and other regression models. *The Annals of Applied Statistics*. 1360-1383.

Gelman, A., & Loken, E. (2013). The garden of forking paths: Why multiple comparisons can be a problem, even when there is no "fishing expedition" or "p-hacking" and the research hypothesis was posited ahead of time. *Department of Statistics, Columbia University*.

---

[6] Or even more complicated variance structures (e.g. a variance structure that varies over a predictor – something like the constant rate effect) – see Lazega & Snijders (2015) for an overview of these models.

Th. Gries, S. (2015). The most under-used statistical method in corpus linguistics: multi-level (and mixed-effects) models. *Corpora*, *10*(1), 95-125.
Hodges, J. S. (2014). *Richly parameterized linear models: additive, time series, and spatial models using random effects*. CRC Press.
----(2016) *Estimates on the boundary in random regressions: Statistical methods research done as science rather than math (work in progress).* Online: http://www.biostat.umn.edu/~hodges/PubH8492/rr_lecture.pdf
Ibrahim, J. G., Zhu, H., Garcia, R. I., & Guo, R. (2011). Fixed and random effects selection in mixed effects models. *Biometrics*, *67*(2), 495-503.
Jaeger, T. F. (2009). Random effects: Should I stay or should I go? URL: https://hlplab.wordpress.com/2009/05/14/random-effect-structure/
Johnson, D. E. (2009). Getting off the GoldVarb standard: Introducing Rbrul for mixed-effects variable rule analysis. *Language and linguistics compass*, *3*(1), 359-383.
Kruschke, J. (2014). *Doing Bayesian data analysis: A tutorial with R, JAGS, and Stan*. Academic Press.
Lazega, E., & Snijders, T. A. (2015). *Multilevel Network Analysis for the Social Sciences*. Springer.
Litière, S., Alonso, A., & Molenberghs, G. (2007). Type I and Type II Error Under Random-Effects Misspecification in Generalized Linear Mixed Models. *Biometrics*, *63*(4), 1038-1044.
McElreath, R. (2016). *Statistical rethinking: A Bayesian course with examples in R and Stan* (Vol. 122). CRC Press.
McCulloch, C. E., & Neuhaus, J. M. (2011). Prediction of random effects in linear and generalized linear models under model misspecification. *Biometrics*, *67*(1), 270-279.
Moineddin, R., Matheson, F. I., & Glazier, R. H. (2007). A simulation study of sample size for multilevel logistic regression models. *BMC medical research methodology*, *7*(1), 1.
Molenberghs, G., & Verbeke, G. (2005). *Models for discrete longitudinal data*. New York: Springer.
Nicenboim, B., & Vasishth, S. (2016) Statistical methods for linguistic research: Foundational Ideas—Part II. *Language and Linguistics Compass*, 10: 591–613. doi: 10.1111/lnc3.12207.
Roy, J. (2006). *The Constant Rate Effect in the Spread of Syntactic Change: Applications of Alternating Logistic Regressions in Historical Repeated Response Data.* M.Sc. Thesis. Statistics, University of Texas at San Antonio.
---- (2009). Inter-dependency in linguistic data from one speaker: Assessing Competing Statistical Models. *38th New Ways of Analyzing Variation*, University of Ottawa. (October 22-25).
----(2012). Sociolinguistic Statistics. In Alena Barysevich, Alexandra D'Arcy and David Heap (eds.), *Proceedings of Methods XIV: Papers from the Fourteenth International Conference on Methods in Dialectology*, pp. 261-75.
----(2014). *The Perfect Approach to Adverbs: Applying variation theory to competing models*. Ph.D. Dissertation. Linguistics, University of Ottawa.
Roy, J and S. Levey. (2014). Mixed Effects Models in Sociolinguistcs: The need for caution. *43$^{rd}$ New Ways of Analyzing Variation*, Chicago (October 22-26)
Ryoo, J. H. (2011). Model selection with the linear mixed model for longitudinal data. *Multivariate behavioral research*, *46*(4), 598-624.

Simmons, J. P., Nelson, L. D., & Simonsohn, U. (2011). False-positive psychology undisclosed flexibility in data collection and analysis allows presenting anything as significant. *Psychological science*, 0956797611417632.
Stan Development Team. 2016. *Stan Modeling Language Users Guide and Reference Manual,* Version 2.14.0.   http://mc-stan.org
Steegen, S., Tuerlinckx, F., Gelman, A., & Vanpaemel, W. (2016). Increasing transparency through a multiverse analysis. *Perspectives on Psychological Science*, *11*(5), 702-712.
Verbeke, G., & Lesaffre, E. (1997). The effect of misspecifying the random-effects distribution in linear mixed models for longitudinal data. *Computational Statistics & Data Analysis*, *23*(4), 541-556.
Wood, S. (2006). *Generalized additive models: an introduction with R*. CRC press.

**Appendix A: Convergence by lme4 default**

| MEM Type | Non-Convergence lme4 | Non-Convergence RStan | Number of Simulations |
|---|---|---|---|
| **Simple Linear** | .008 % | .009 % | 80,000 |
| **Simple Logistic** | 22 % | .002 % | 20,000 |
| **Complex Linear** | 14 % | 3 % | 2,500 |
| **Complex Logistic** | 98 % | <.001 % | 2,500 |

The table above has the non-convergence rates if we set the tolerance = .002 (the lme4 default).

**Appendix B: Full Generalized Additive Model Results for Section 3.**

Imbalance, Minimum Variance, Sample Size and Convergence.

```
Formula:
conv ~ fam + s(balance) + s(minvar) + s(S, k = 5)

Parametric coefficients:
            Estimate Std. Error z value Pr(>|z|)
(Intercept)  1.91432    0.06025   31.77   <2e-16 ***
famLogistic -3.54700    0.08458  -41.94   <2e-16 ***
---

Approximate significance of smooth terms:
            edf Ref.df  Chi.sq  p-value
s(balance) 6.071  7.253 117.994  < 2e-16 ***
s(minvar)  2.230  2.819  40.301 1.84e-08 ***
s(S)       1.303  1.541   1.219    0.546
---

R-sq.(adj) =  0.481   Deviance explained = 39.2%
UBRE = -0.15362  Scale est. = 1         n = 5000
```

Accuracy of Fixed Effects in Converged Models

Family: gaussian
Link function: identity

Formula:
error ~ type * fam * reg + s(balance)

Parametric coefficients:

|  | Estimate | Std. Error | t value | Pr(>\|t\|) |  |
|---|---|---|---|---|---|
| (Intercept) | 0.0648865 | 0.0048381 | 13.411 | < 2e-16 | *** |
| typeb1_error | -0.0288368 | 0.0068419 | -4.215 | 2.50e-05 | *** |
| typeb2_error | -0.0204869 | 0.0068419 | -2.994 | 0.002751 | ** |
| typeb3_error | -0.0197422 | 0.0068419 | -2.885 | 0.003909 | ** |
| typer01_error | 0.1324452 | 0.0068419 | 19.358 | < 2e-16 | *** |
| typer02_error | 0.1679870 | 0.0068419 | 24.553 | < 2e-16 | *** |
| typer03_error | 0.1623082 | 0.0068419 | 23.723 | < 2e-16 | *** |
| typer12_error | 0.1895573 | 0.0068419 | 27.705 | < 2e-16 | *** |
| typer13_error | 0.1941156 | 0.0068419 | 28.371 | < 2e-16 | *** |
| typer23_error | 0.2059077 | 0.0068419 | 30.095 | < 2e-16 | *** |
| types0_error | -0.0110786 | 0.0068419 | -1.619 | 0.105405 |  |
| types1_error | -0.0335315 | 0.0068419 | -4.901 | 9.56e-07 | *** |
| types2_error | -0.0214142 | 0.0068419 | -3.130 | 0.001750 | ** |
| types3_error | -0.0232463 | 0.0068419 | -3.398 | 0.000680 | *** |
| famLogistic | 0.2962872 | 0.0116820 | 25.363 | < 2e-16 | *** |
| regStan | -0.0015345 | 0.0066387 | -0.231 | 0.817205 |  |
| typeb1_error:famLogistic | -0.1501126 | 0.0165158 | -9.089 | < 2e-16 | *** |
| typeb2_error:famLogistic | -0.0665941 | 0.0165158 | -4.032 | 5.53e-05 | *** |
| typeb3_error:famLogistic | -0.0284402 | 0.0165158 | -1.722 | 0.085074 | . |
| typer01_error:famLogistic | 0.0409914 | 0.0165158 | 2.482 | 0.013068 | * |
| typer02_error:famLogistic | 0.0186057 | 0.0165158 | 1.127 | 0.259942 |  |
| typer03_error:famLogistic | 0.0479491 | 0.0165158 | 2.903 | 0.003694 | ** |
| typer12_error:famLogistic | 0.0104404 | 0.0165158 | 0.632 | 0.527294 |  |
| typer13_error:famLogistic | 0.1019103 | 0.0165158 | 6.170 | 6.83e-10 | *** |
| typer23_error:famLogistic | 0.0846158 | 0.0165158 | 5.123 | 3.01e-07 | *** |
| types0_error:famLogistic | -0.1641078 | 0.0165158 | -9.936 | < 2e-16 | *** |
| types1_error:famLogistic | -0.2052171 | 0.0165158 | -12.425 | < 2e-16 | *** |

```
types2_error:famLogistic                -0.1022181  0.0165158  -6.189 6.07e-10 ***
types3_error:famLogistic                -0.1225240  0.0165158  -7.419 1.19e-13 ***
typeb1_error:regStan                     0.0009412  0.0093883   0.100 0.920148
typeb2_error:regStan                     0.0008980  0.0093883   0.096 0.923802
typeb3_error:regStan                     0.0007132  0.0093883   0.076 0.939444
typer01_error:regStan                   -0.0210357  0.0093883  -2.241 0.025053 *
typer02_error:regStan                   -0.0315496  0.0093883  -3.361 0.000778 ***
typer03_error:regStan                   -0.0333225  0.0093883  -3.549 0.000386 ***
typer12_error:regStan                   -0.0355720  0.0093883  -3.789 0.000151 ***
typer13_error:regStan                   -0.0477078  0.0093883  -5.082 3.75e-07 ***
typer23_error:regStan                   -0.0382935  0.0093883  -4.079 4.53e-05 ***
types0_error:regStan                    -0.0039028  0.0093883  -0.416 0.677623
types1_error:regStan                     0.0001350  0.0093883   0.014 0.988529
types2_error:regStan                    -0.0031816  0.0093883  -0.339 0.734696
types3_error:regStan                    -0.0020526  0.0093883  -0.219 0.826936
famLogistic:regStan                     -0.1876487  0.0133129 -14.095  < 2e-16 ***
typeb1_error:famLogistic:regStan         0.1159846  0.0188239   6.162 7.23e-10 ***
typeb2_error:famLogistic:regStan         0.0775377  0.0188239   4.119 3.81e-05 ***
typeb3_error:famLogistic:regStan         0.0415666  0.0188239   2.208 0.027234 *
typer01_error:famLogistic:regStan        0.0038370  0.0188239   0.204 0.838481
typer02_error:famLogistic:regStan        0.0112713  0.0188239   0.599 0.549324
typer03_error:famLogistic:regStan       -0.0069289  0.0188239  -0.368 0.712806
typer12_error:famLogistic:regStan        0.0174507  0.0188239   0.927 0.353901
typer13_error:famLogistic:regStan       -0.0697000  0.0188239  -3.703 0.000213 ***
typer23_error:famLogistic:regStan       -0.0619876  0.0188239  -3.293 0.000991 ***
types0_error:famLogistic:regStan         0.1342552  0.0188239   7.132 9.94e-13 ***
types1_error:famLogistic:regStan         0.1868890  0.0188239   9.928  < 2e-16 ***
types2_error:famLogistic:regStan         0.1047386  0.0188239   5.564 2.64e-08 ***
types3_error:famLogistic:regStan         0.1276494  0.0188239   6.781 1.20e-11 ***
---
Approximate significance of smooth terms:
            edf Ref.df     F p-value
s(balance) 8.257  8.852 214.7  <2e-16 ***
---
R-sq.(adj) =  0.247   Deviance explained = 24.7%
GCV = 0.050307  Scale est. = 0.050276  n = 105350
```